\documentclass{optica-article}

\journal{opticajournal} 

\articletype{Research Article}

\usepackage{lineno}

\usepackage{dcolumn}
\usepackage{bm}
\usepackage{makecell}
\usepackage{svg}
\usepackage{amsmath}
\usepackage{float}
\usepackage{multirow, tabularray}
\usepackage[normalem]{ulem}
\useunder{\uline}{\ul}{}

\newcommand{\ket}[1]{\ensuremath{|{#1}\rangle}}

\newcommand{\mic}{\,$\mu$m\:}

\newcommand{\siox}{SiO$_2$\:}
\newcommand{\sini}{SiN$_x$\:}

\begin{document}

\title{Integrated photonic structures for photon-mediated entanglement of trapped ions}

\author{F. W. Knollmann,\authormark{1,*} E.~Clements,\authormark{1} P. T. Callahan,\authormark{2} M.~Gehl,\authormark{3} J. D.~Hunker,\authormark{3} T.~Mahony,\authormark{2} R.~McConnell,\authormark{2} R.~Swint,\authormark{2} C. Sorace-Agaskar,\authormark{2} I. L.~Chuang,\authormark{1} J. Chiaverini,\authormark{1,2} and D. Stick\authormark{3}}

\address{\authormark{1}Massachusetts Institute of Technology, Cambridge, Massachusetts 02139, USA\\
\authormark{2}Lincoln Laboratory, Massachusetts Institute of Technology, Lexington, Massachusetts 02420, USA\\
\authormark{3}Sandia National Laboratories, Albuquerque, New Mexico 87185, USA}

\email{\authormark{*}fwk@mit.edu} 


\begin{abstract} 
Trapped atomic ions are natural candidates for quantum information processing and have the potential to realize or improve quantum computing, sensing, and networking.  
These applications often require the collection of individual photons emitted from ions into guided optical modes, in some cases for the production of entanglement between separated ions.
Proof-of-principle demonstrations of such photon collection from trapped ions have been performed using high-numerical-aperture lenses or cavities and single-mode fibers, but integrated photonic elements in ion-trap structures offer advantages in scalability and manufacturabilty over traditional optics.
In this paper we analyze structures monolithically fabricated with an ion trap for collecting ion-emitted photons, coupling them into waveguides, and manipulating them via interference. 
We calculate geometric limitations on collection efficiency for this scheme, simulate a single-layer grating that shows performance comparable to demonstrated free-space optics, and discuss practical fabrication and fidelity considerations. Based on this analysis, we conclude that integrated photonics can support scalable systems of trapped-ions that can distribute quantum information via photon-mediated entanglement.
\end{abstract}

\section{Introduction}
\label{sec:intro}

Systems of trapped atomic ions are among the best prospects for practical quantum computing, as ions of the same species form identical qubits, are well isolated from noise sources, and allow high-fidelity laser- or microwave-based quantum-logic operations~\cite{wineland:1998,bcms_2019}. Moreover, their natural ability to interface with photonic flying qubits  makes ions a promising technology for applications that require both long memory times and photonic qubit transduction, including quantum networks~\cite{drmota:2023}, quantum repeaters~\cite{muralidharan:2015}, and quantum sensing~\cite{komar2014quantum}.
These applications rely on coupling photons emitted from individual ions into single mode structures to mediate entanglement between separate nodes. This procedure is often called remote entanglement generation because it supports large distances between nodes, as in a quantum repeater.  More generally, it is known as {\em photon mediated entanglement} (PME), for it can also connect separate but nearby modules in a single quantum computer~\cite{brown:2016} or a more generic quantum information processing system. 

The traditional technique used in trapped-ion-based PME experiments relies on a high numerical aperture (NA) lens to collect light into a single mode fiber. Early experiments demonstrated 0.4\% collection efficiency $\eta$ of photons from each of two ions into fibers to achieve a remote entanglement rate $r_{\rm ent}=2$ mHz~\cite{moehring:2007}.  Since $r_{\rm ent}$ scales with $\eta^2$, considerable gains are possible by increasing the collection efficiency using large lens elements positioned in close proximity to the emitters or by changing the vacuum mode structure through use of a cavity~\cite{krutyanskiy:2023}. Using the former strategy, recent experiments have demonstrated $\eta \approx$ 3.5\% and $r_{\rm ent}=182$ Hz~\cite{ballance:2020}. These experiments involved entangling each ion qubit with photon states distinguished by their polarization or frequency, though other schemes using the photon number or arrival time are possible.

Recent improvements for PME notwithstanding, these traditional approaches will not easily scale to larger systems due mainly to the limitations of free-space optics for imaging extended arrays of closely spaced ions~\cite{moody:2022}.  Modular architectures will likely require communication regions with PME capabilities evenly interspersed amongst qubit ions, rather than located in a central region \cite{monroe:2014}.  The size of high-numerical-aperture bulk optics is incommensurate with the attainable array pitch since traditional lenses consume considerably more lateral chip area than they can efficiently image.  At one extreme is a single-element lens, which is excellent at imaging a single site but degrades rapidly off-axis. In \cite{carter:2023}, two single-element lenses above and below the trap each collect 5\% of the emitted light into separate fibers. A Zemax simulation of this lens using the Huygens point-spread-function shows that an ion that is 10\mic off-axis couples a factor of -10.3 dB fewer photons into a single mode fiber at the appropriate location compared to the ion that is on-axis.  A multi-lens system can perform better for off-axis ions, but is much larger and may be difficult to integrate into a system due to size constraints.  For instance, the main imaging lens in \cite{clark:2021} has a 0.63 NA and is 50 mm in diameter with a 160\mic diameter field of view.  While this field of view could accommodate the $\approx$50 ions that could fit in a 160\mic chain, it would still consume considerably greater lateral area than its effective PME collection area.  This is especially true for an architecture in which ions are spaced farther apart, like a QCCD array with a $\sim$250\mic pitch.  At the other extreme are lens systems like those used for photolithography, which have space-bandwidth products of thousands of gigapixels \cite{zheng:2021} and may be able to collect photons for PME from a large ion array, but are bulky and likely difficult to integrate with the rest of a trapped-ion quantum information processing system.  It is possible to use ion transport to shuttle ions from the periphery of an array to a central collection zone, but this would limit parallel PME operation and incur time costs due to shuttling.

Alternatives that are more scalable include miniaturized optics that are hybrid or monolithically integrated with the trap to collect photons emitted from ions into single-mode waveguides or fibers. Examples of hybrid integration of separate optical structures with ion traps include microlens arrays~\cite{birkl:2001}, micro-cavities~\cite{kim:2011}, and diffractive elements or metamaterials on separate substrates~\cite{clark:2014,ropp:2023}.
Separating the optical elements from the trap simplifies fabrication and eases trap geometry constraints. 
However, these approaches face significant alignment challenges, since maximizing photon collection efficiency is accompanied by tight positional tolerances.

Monolithically integrated collection optics have a significant advantage because they can be precisely aligned using lithographic techniques. Diffractive reflectors have been demonstrated for photon collection into single free-space modes~\cite{ghadimi:2017,connell:2021}, but on-chip waveguide optics offer additional capabilities for photon manipulation.  We therefore focus on collection elements based on trap-integrated diffractive waveguide-beam couplers, also known as grating couplers, such as those that have been employed for directing light to ions~\cite{mehta:2016, niffenegger:2020, mehta:2020, ivory:2021,PhysRevLett.130.133201}.
These demonstrations have shown improvements in beam-pointing and phase stability as well as resilience to vibrations of the ion-trap system~\cite{10.1063/1.3058605,10.1063/1.4802948,Pagano_2019}. Beyond scalability, integrated photonic elements based on this technology may also offer similar improvements to the rate and fidelity of PME.  While we do not consider cavity integration in this manuscript due to the significant additional fabrication challenges associated with resonant structures, we note that the use of resonant cavities for light collection has led to efficiencies as high as $57\%$ \cite{Schupp2021}.

In this paper we describe and analyze schemes for monolithically integrating diffractive collection optics to scalably support photon-mediated entanglement in trapped-ion systems.  As part of this analysis, we calculate performance limitations inherent to surface ion traps and simulate the optical performance for a simple prototype optic.  We further analyze several other integrated optical components needed to support PME, as well as their assembly into photonic integrated circuits (PICs). This analysis is differentiated among relevant PME protocols that use photon interference and erasure of which-path information to create heralded Bell pairs.  We show how these protocols imply particular benefits and limitations when using waveguide-based integrated optical elements, and we suggest avenues to improve the PME rate by leveraging optical integration to perform parallel operations over many ion pairs.  

We begin by briefly describing relevant protocols for PME with an emphasis on requirements when using integrated diffractive optics for collection in Sec.~\ref{sec:schemes}, then lay out integrated optical components with the potential to realize these protocols in Sec.~\ref{sec:components}.  We describe implications of their use in Sec.~\ref{sec:analysis}.

\section{Protocols for photon mediated entanglement}
\label{sec:schemes}

\begin{figure}[htbp]
\centering
\includegraphics[width=0.45\textwidth]{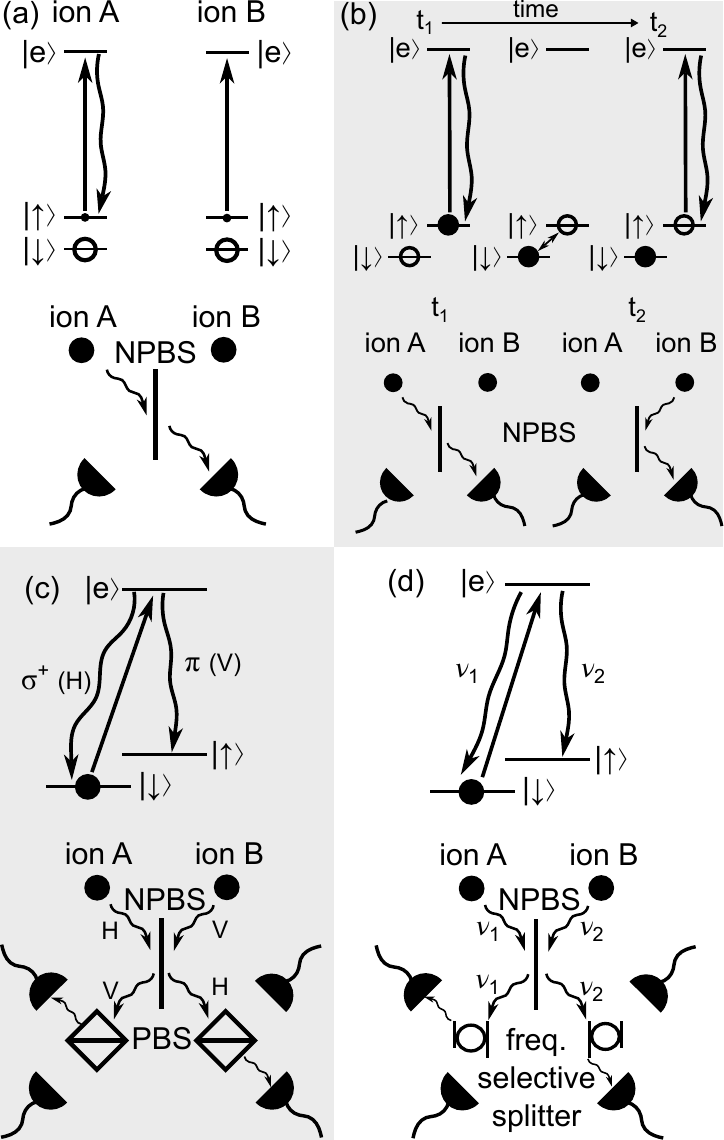}
\caption{Photon mediated entanglement protocols using four different photonic variables: number (a), time-bin (b), polarization (c), and frequency (d). We show an improved Bell-state analyzer to double the entanglement rate for the polarization and frequency protocols \cite{simon:2003}. Note that discriminating between the two photons based on frequency is nontrivial for typical gigahertz splittings.  PBS:  Polarizing beamsplitter.  NPBS:  Non-polarizing beamsplitter.
}
\label{fig:schemes}
\end{figure}

\begin{table*}[htbp]
\begin{tblr}{colspec  = {X[.4,c,m]X[2.6,r,m]|[1pt,gray]X[1.8,c,m]X[1.8,c,m]X[2,c,m]X[1.8,c,m]},
            colsep=4pt,
            row{1}   = {bg=gray8, font=\bfseries}, 
            row{even} = {bg=gray8!30},    
            column{1} = {bg=white, font=\large\bfseries},
            column{2} = {font=\bfseries},
            rowsep=1pt
            }
    & Photonic variable & Number & Time-bin & Polarization & Frequency \\
    \cline{1-6}
    \SetCell[r=5]{bg=gray9}{\rotatebox[origin=c]{90}{Atomic}}
    & Initial state for pulsed excitation &
        superposition state &
        superposition state &
        projected state & 
        projected state\\
    & Control pulses &
        SP, (1Q), PE &    
        SP, 1Q, PE & 
        SP, PE &    
        SP, PE \\
    & Probability of successful herald &
        $2 p_e \gamma \epsilon$ & 
        $\frac{1}{2}(p_e \gamma \epsilon)^2$ &    
        $\frac{1}{2}(p_e \gamma \epsilon)^2$ &  
        $\frac{1}{2}(p_e \gamma \epsilon)^2$ \\
    & Main fidelity limitations &
        double excitation, ion temperature, loss imbalance &   
        single qubit rotations &   
        mixing of polarization states in PIC &  
        mixing of frequency states \\
    & Requirement for path length stability &
        wavelength-scale &  
        cm-scale &      
        cm-scale & 
        cm-scale \\
    \cline{1-6}
    \SetCell[r=2]{bg=gray8}{\rotatebox[origin=c]{90}{Photonic}}
    & Integrated components &
        collection gratings, splitter, detector & 
        collection gratings, splitter, detector &      
        polarization selective collection gratings, mode-dependent splitter/combiner, mode-agnostic splitter, detector &   
        collection gratings, splitter, frequency dependent splitter, detector \\
    & Fidelity depends on collection angle & 
        no &
        rate only
        no &
        yes &
        no  \\ 
    \cline{1-6}
\end{tblr}
\caption{A comparison of four schemes (number, time-bin, polarization, and frequency) used for generating PME, specifying constraints and requirements on the atomic and photonic subsystems involved. The analysis assumes a ground state qubit encoded in levels $\ket{\uparrow}$ and $\ket{\downarrow}$. Each protocol begins with a form of state preparation (SP) into a definite quantum state.  Some require interleaved single-qubit rotations (1Q) and all rely on pulsed excitation (PE) from the ground state manifold to a fast decaying excited level with probability $p_e$. Note that the number scheme \textit{may} require a single-qubit rotation, depending on the implementation. The photon used for heralding entanglement is generated by spontaneous decay from this excited level back to the ground state manifold with probability given by the branching ratio $\gamma$. The probability of a detector click caused by this photon is $\epsilon = \frac{\Omega}{4\pi} p_t \eta_D$. This probability is split into the collection probability given by the detection solid angle $\Omega$ (we ignore the radiation emission profile of the ion for simplicity here), the transmission probability from collection optic to detector $p_t$ and the detector quantum efficiency $\eta_D$. There is a numerical factor related to the fraction of possible Bell states that generate viable heralding patterns. Here we assume the maximum efficiency achievable for linear optics and non-number-resolving detectors for the polarization and frequency protocols.}
\label{tab:schemes}
\end{table*}

Photon mediated entanglement between two ions is typically accomplished with the ions each encoding a qubit in their ground electronic states, labelled $\ket{\uparrow}$, $\ket{\downarrow}$, which can be coupled via photons to an excited state $| e \rangle$. Entanglement between the two ions proceeds by initial ion-photon entanglement in each ion of a pair, the subsequent heralded transfer of that entanglement to the two ions via the erasure of which-path information through photon interference (e.g.\ the Hong-Ou-Mandel effect~\cite{HOM_1987}), and photon detection. Certain detection patterns project the ions into an entangled state, generally one of the Bell states $| \Psi^{\pm} \rangle = (|\! \uparrow_1  \downarrow_2 \rangle \pm |\! \downarrow_1  \uparrow_2 \rangle / \sqrt{2}$, conditioned on the measurement result.  Several schemes for this process have been devised, with the photonic variables used as intermediaries being the most natural way to categorize them.  These are briefly described here (see Fig.~\ref{fig:schemes}), with particular attention paid to their realizations with PICs and corresponding relative strengths and weaknesses (see Table~\ref{tab:schemes}). The details of these protocols and their implementation with other architectures are beyond the scope of this paper and we point the interested reader to the relevant references~\cite{feng:2003,duan:2003,simon:2003,cabrillo:1999,Luo_2009}.

Each of these schemes relies on the photons from each ion being indistinguishable, both temporally and spatially, for high fidelity erasure of which-path information. Mode matching in the temporal dimension is determined by the timing of the excitation pulse, the decay profile of the atomic transition used, and the path length difference between each ion and the beamsplitter. The spatial mode matching is determined at the beamsplitter. Integrated photonic components are in many cases more capable of meeting these requirements than free-space optics, as we describe below and in Sec.~\ref{sec:analysis}.

The PME fidelity $F$ is given by the overlap of the actual state that is generated experimentally, $|\psi \rangle$, with the ideal Bell state $| \Psi^{\pm} \rangle$: $F = | \langle \Psi^{\pm} | \psi \rangle |^2 $.  In some cases, such as for poor temporal overlap, one can trade-off entanglement rate and fidelity by changing the post-selection parameters (i.e. by narrowing the acceptance time window for a valid herald pattern~\cite{krutyanskiy:2023}).
The required entanglement fidelity is ultimately governed by application requirements. In the case of quantum computing, one generally desires low-overhead quantum error correction codes, which generally have acceptable infidelities $<$1\%.
The PME fidelity does not have to reach this threshold because entanglement distillation can increase the quality of entanglement at the cost of a proportional overhead of lower quality entangled pairs~\cite{nickerson:2014,bennet_1996}.   However, to the extent that additional overhead can greatly increase the resources or time needed to perform a computation, maximizing entanglement fidelity (while still maintaining a reasonable $r_{\rm{ent}}$) is highly desirable.

\subsection{Number-based entanglement}

The simplest protocol for PME relies on exciting each emitter with low probability $p_{e}$, followed by detection of a single photon to project one (and ideally only one) ion into the excited state; the simultaneous emission of photons from both ions is unlikely, with a probability $p_{e}^{2}$ \cite{cabrillo:1999}.  This method has been demonstrated using trapped ions~\cite{slodicka:2013} and nitrogen vacancy (NV) centers in diamond~\cite{humphreys:2018}. As shown in Fig.~\ref{fig:schemes}(a), each ion is prepared in a superposition state $\sqrt{(1-p_e)}\ket{\downarrow}+\sqrt{p_e}\ket{\uparrow}$, where $p_e\ll1$, either via a partial single-qubit rotation followed by unit-probability excitation on a cycling transition or by weak optical excitation from $\ket{\downarrow}\rightarrow\ket{e}$. The state \ket{e} then decays back to \ket{\uparrow} via spontaneous emission of a photon. Selection rules or polarization/frequency filtering are used to guarantee no decay from $\ket{e}$ to $\ket{\downarrow}$. The photon decay pathways from both ions are combined on a beamsplitter and measured using two detectors as shown in Fig.~\ref{fig:schemes}(a). If a single count is registered across both detectors, then with high probability the lack of which-path information results in an entangled state with only one ion having been transferred to \ket{\uparrow}. Due to the low probability of excitation and assuming the overall detection efficiency is small, the resulting ion state to lowest-order in $p_e$ is:
\begin{equation}
\ket{\psi} = \sqrt{\frac{(1-p_e)}{2}} (\ket{\uparrow_1 \downarrow_2}\pm e^{i \phi}\ket{\downarrow_1 \uparrow_2})+\sqrt{p_e}\ket{\uparrow_1 \uparrow_2},
\label{eq:number}
\end{equation}
where the first term can be converted to a Bell state by local operations, $\phi$ depends on the difference in optical path length from each ion to the beamsplitter ($k \Delta l$), and the sign depends on which detector registered a click. 

This protocol requires a trade-off between entanglement rate and fidelity. On the one hand, $r_{\rm ent}$ increases proportionally with $p_e$, since the success probability scales as $p_{\rm{success}} = 2 p_e \gamma \epsilon$, where $\gamma$ is the branching ratio and the probability of a detector click with ion emission is $\epsilon = \frac{\Omega}{4\pi} p_t \eta_D$, with collection solid angle $\Omega$, photon transmission probability $p_t$, and detector quantum efficiency $\eta_D$. On the other hand, the excitation probability $p_e$ must be kept low to minimize the probability of both ions emitting a photon, since double-excitation limits the fidelity to $(1-p_e)$ under the assumption of low overall detection efficiencies. 

There are three further fidelity limitations specific to this protocol. First is that the optical path length difference must remain stable as it determines the relative phase factor of the final state. This requirement should be eased by the inherently better interferometric stability of integrated photonic circuits. Second, the combined probability of exciting and subsequently detecting a photon from either ion must be balanced. Any imbalance leads to partial knowledge of which ion is in state $\ket{\uparrow}$~\cite{hermans:2023}. Finally, the fidelity is limited by the ability to determine which ion scattered a photon because of motional excitation due to photon recoil~\cite{cabrillo:1999,slodicka:2013}. The infidelity grows strongly with increased average phonon number and can be mitigated by ground-state cooling or tighter trapping~\cite{Luo_2009}.

\begin{table*}[htbp]
\begin{tblr}{colspec  = {X[0.5,c,m]|[1pt,gray]X[1,c,m]|[1pt,gray]X[1,c,m]|[1pt,gray]X[2.5,c,m]},
            colsep=4pt,
            row{1}   = {bg=gray8, font=\bfseries}, 
            row{even} = {bg=gray8!30},      
            column{1} = {font=\bfseries},
            rowsep=1pt
            }
    Ion & P$\mathbf{_{1/2}}$ $\mathbf{\lambda}$ (nm) & P$\mathbf{_{3/2}}$ $\mathbf{\lambda}$ (nm) & \SetCell[r=1]{}{Ground st. hyperfine \\(A: splitting [GHz])}\\
    \cline{1-4}
     Ca$^+$ & 397 & 393 & 43: 3.2\\
    \cline{1-4}
    Sr$^+$ & 422 & 408 & 87: 5.0\\
    \cline{1-4}
    Ba$^+$ & 493 & 455 & \SetCell[r=1]{}{133: 9.9\\137: 8.0}\\
    \cline{1-4}
    Yb$^+$ & 369 & 329 & \SetCell[r=1]{}{171: 12.6\\ 173: 10.5}\\
    \cline{1-4}
\end{tblr}
\caption{Common ion species and their transition wavelengths most suitable for given protocols. The S$_{1/2}\rightarrow$P$_{1/2}$ transitions are favorable for the polarization protocol, while S$_{1/2}\rightarrow$P$_{3/2}$ transitions can be favorable for number and time-bin entanglement because of the available closed transition. Note that the frequency protocol works for either transition. However, it requires the larger ground-state frequency splitting of ions with hyperfine structure. Representative isotopes and their ground-state hyperfine splitting are shown in the last column (values from~\cite{sunaoshi:1993,knab:2007,blatt:1982,munch:1987}).
The other protocols are more straightforwardly implemented on ions without hyperfine structure. An exception is the polarization protocol as described for $^{171}$Yb$^+$ in~\cite{Luo_2009}. We have omitted decay channels with transition wavelengths in the IR since their low branching ratios lead to very low rates without cavity enhancement~\cite{krutyanskiy:2023}.
}
\label{tab:species}
\end{table*}

\subsection{Time-bin entanglement}

A second scheme relies on detecting photon emission in two distinct time bins to project the ions into an entangled state. This protocol has primarily been demonstrated using NV centers~\cite{tchebotareva:2019}. In time-bin entanglement, as depicted in Fig.~\ref{fig:schemes}(b), only one of the ion ground states (\ket{\uparrow} in the figure) couples to \ket{e}. We assume here that both ions begin in an equal superposition of \ket{\uparrow} and \ket{\downarrow}. An optical pulse applied to both ions couples $\ket{\uparrow} \leftrightarrow \ket{e}$, resulting in an entangled state where the presence (absence) of an emitted photon corresponds to the ion in \ket{\uparrow} (\ket{\downarrow}). Next a $\pi$ pulse applied to both ions inverts the populations of the ground states. A final optical pulse applied to both ions again couples $\ket{\uparrow} \leftrightarrow \ket{e}$ leading to a similar ion-photon entangled state as with the first pulse. The two optical pulses must be separated in time by more than the lifetime of \ket{e} and the recovery time of the detector. Entanglement is heralded by collecting and detecting a photon after both optical pulses. This implies that one of the two ions projected to \ket{\uparrow} after the first optical pulse was then transferred to \ket{\downarrow} by the $\pi$ pulse, and then the second ion was projected to \ket{\uparrow} after the second optical pulse. Thus, observing photons in both time bins results in the entangled state:
\begin{equation}
\ket{\psi_\pm} = \frac{1}{\sqrt{2}} (\ket{\uparrow_1 \downarrow_2}\pm e^{i \phi}{\ket{\downarrow_1 \uparrow_2}}),
\label{TimeEntangleState}
\end{equation}
where $\phi =\Delta k \Delta l + \Delta \omega \Delta t$ and the sign depends on whether the same detector or different detectors clicked in the two time-bins. The phase depends on the optical path length difference between the two arms times the wavenumber difference between the two photons, which is small, since $\Delta k \ll k$. It also depends on the time between pulses multiplied by the qubit frequency difference ($\ket{\downarrow}\rightarrow \ket{\uparrow}$) between ions. The $\Delta \omega \Delta t$ term is generally small, but shows that the qubit frequency difference should be stable over the course of the experiment. Overall, $\phi$ is constant across experiments to good approximation and this state can thus be reliably turned into a Bell state by single qubit rotations.

The probability of a successful herald for this and the other multi-photon entanglement schemes scales as $\frac{1}{2}(p_e \gamma \epsilon)^2$ because two photons must be collected and detected. However, the excitation probability $p_e$ should approach unity, so the heralding probability can be comparable to the number-based entanglement case. Furthermore, path length stability is required only to the wavelength of the difference frequency between the two photons $2 \pi c /\Delta \nu$, which is typically centimeter-scale or greater and is thus very relaxed compared to the number-based entanglement protocol \cite{simon:2003}.

\subsection{Polarization entanglement}

Polarization-based entanglement takes advantage of excited-state decay into two orthogonal polarization states. For this scheme, as depicted in Fig.~\ref{fig:schemes}(c), the excited state \ket{e} can decay into either of the possible ground states, \ket{\uparrow} and \ket{\downarrow}, via emission of photons with different angular momentum, such as \ket{\pi} and \ket{\sigma^+}, which map to different polarizations when measured along specific axes relative to the quantization axis set by the applied static magnetic field. This protocol is also possible with other photon polarizations, e.g.\ \ket{\sigma^-} and \ket{\sigma^+}, as long as the polarization states used are orthogonal at the point of interference to avoid degraded fidelity of the entangled state~\cite{ballance:2020}. Each ion is maximally entangled with the photon it emits. The decay photons are collected from both ions and interfered using a non-polarizing beamsplitter (NPBS).
Due to photon statistics, a measurement of one photon in each output port of the NPBS projects the ions into the corresponding anti-symmetric Bell state $\ket{\Psi^-} =1/\sqrt{2} \left( \ket{\downarrow_1 \uparrow_2} - \ket{\uparrow_1 \downarrow_2} \right)$ up to a local relative phase factor $e^{i\phi}$. Adding PBSs to the Bell state analyzer can improve $r_{\rm ent}$ by a factor of 2 by additionally identifying the symmetric Bell state $\ket{\Psi^+} = 1/\sqrt{2} \left( \ket{\downarrow_1 \uparrow_2} + \ket{\uparrow_1 \downarrow_2}\right)$, where both photons exit the same NPBS output with orthogonal polarizations and can thus be split by the PBS and detected separately.  This state can be transformed to the anti-symmetric Bell state, and the local phase $\phi = \Delta k \Delta l$ can be compensated, through a single-ion rotation.

\subsection{Frequency-based entanglement}

Frequency-based entanglement, as depicted in Fig.~\ref{fig:schemes}(d), proceeds analogously to polarization entanglement but with the ion state information encoded in the photon frequency rather than its polarization.  The photon frequency difference between decay channels must be sufficient to make the photons distinguishable and preclude interference. To improve the Bell state analyzer, the PBSs that are used in polarization-based entanglement must be replaced by frequency-selective elements, such as interferometers. Because hyperfine ground states are typically separated by only gigahertz frequencies (see Table~\ref{tab:species}), these interferometric structures may be relatively large (compared to other PIC components) and potentially challenging to incorporate into an integrated platform.

\begin{figure}[htbp]
\includegraphics[width=1\textwidth]{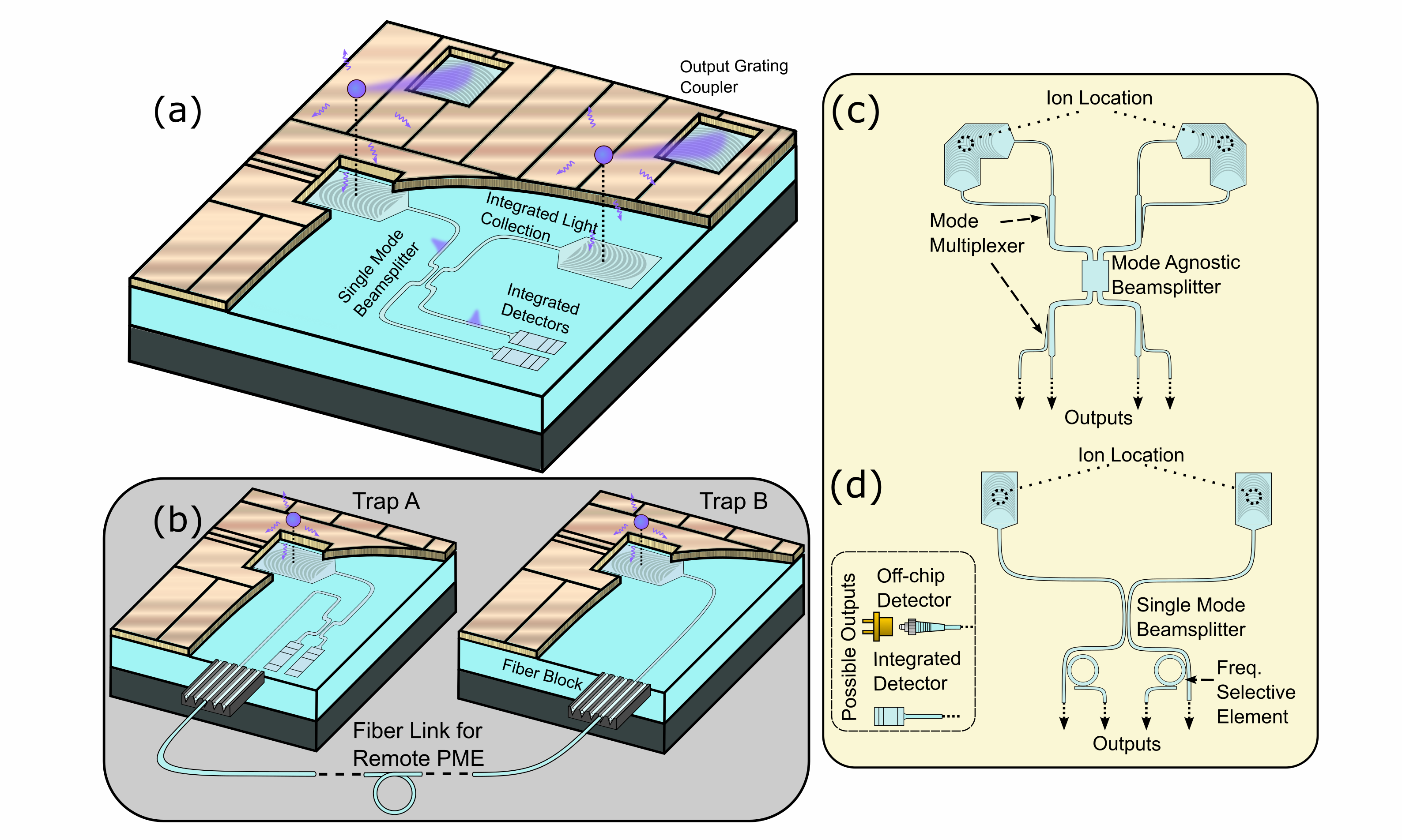}
\caption{Depiction of PICs and components for PME based on monolithic integration in a surface-electrode ion-trap chip. (a) Cut-away illustration containing multiple layers of an ion-trap chip with a PIC embodying the concept for same-chip PME: substrate (black), \siox cladding (blue), waveguide layer with representative photonic devices (gray), and metal layer for trap electrodes (gold).  The PIC layout shown here is compatible with number, time-bin, and frequency (without enhanced Bell-state analysis) based PME. Both emission gratings for ion control (top, not discussed here), and collection gratings for photon collection (center), are shown. (b) Potential layout for remote entanglement generation with collection optics for ions A and B on separate chips with a fiber-based photonic interconnect.  Specific photonic layouts for polarization-based and frequency-based (with enhanced Bell-state analysis via frequency selection) PME can be seen in (c) and (d) respectively. Inset in (d) shows possible options for single-photon detection, either via detector integration or waveguide-to-fiber coupling.}
\label{fig:overview}
\end{figure}

\section{Integrated Optics for Photon Collection and Interference}
\label{sec:components}

The protocols for photon mediated entanglement map to different photonic integrated circuit architectures, each with unique requirements for PIC components and component performance. However, they share a general methodology of collecting photons with diffraction gratings into on-chip waveguides, interfering/combining them with beamsplitters, and detecting them with integrated or off-chip detectors. Fig.~\ref{fig:overview} shows a conceptual illustration of what PICs for PME integrated into an ion trap might look like.  Part (a) shows trap integrated elements in a PIC for the number or time-bin protocols, in this case for ions on a single chip, while part (b) depicts chip-to-chip remote entanglement generation. Fig.~\ref{fig:overview}(c) and Fig.~\ref{fig:overview}(d) show circuit diagram examples analogous to that in (a) for polarization based PME and frequency based PME, respectively. 

For photon-number or time-bin based PME the photons from both ions can simply be combined on a waveguide beamsplitter before being sent to the detectors. In contrast, polarization based PME requires multiple photon encodings.  One method to accomplish this is to collect light from two orthogonally polarized decay channels using a combined grating coupler, which couples the two photon polarizations into separate waveguides.  The polarization degree of freedom is mapped to a path degree of freedom. The waveguides deliver the light into a mode converter which maps the two photon paths, and hence the two polarizations, to a fundamental spatial mode and to a higher order spatial mode that then propagate identically through the system in a multimode waveguide. Which-path information is removed using a mode agnostic beamsplitter which interferes both modes independently and either simultaneously separates the two modes or sends them to a mode de-multiplexer for an enhanced Bell-state analyzer. In contrast to the polarization case, frequency based PME requires simpler photonic devices because even relatively narrow bandwidth PIC components tend to have bandwidths of a few nanometers, while the frequency splitting of a photonic qubit produced from an ion is typically ${\sim}$10~GHz (see Table~\ref{tab:species}). However, the simplicity of the photonic elements in frequency PME holds only when a single Bell state can be detected. Enhancing the Bell-state analyzer to increase the rate by a factor of 2 can be challenging due to this small frequency difference and likely requires a resonant device (see further discussion below in Sec.~\ref{sec:freq_components}).   

To maximize PME fidelity and rate, each component in the optical path needs to be designed to optimize performance and minimize loss. Below, we step through the components of the PICs in order and discuss key considerations, performance metrics, and the current state-of-the-art (cf. also Table~\ref{tab:schemes}). 

\subsection{Photon collection gratings}
\label{sec:gratings}
\begin{figure}[!htbp] 
\centering
\includegraphics[width=0.48\textwidth]{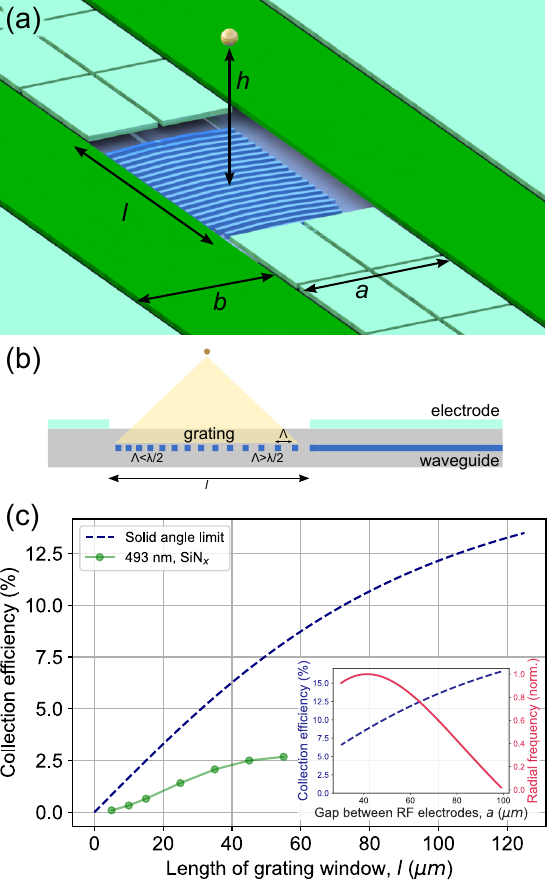}
\caption{Images and simulations of monolithically fabricated collection gratings integrated into an ion trap.  Part (a) shows a conceptual rendering labeled with the dimensions of the RF electrodes (dark green, both electrodes of the same width), control electrodes (light green), and an ion that is centered above the grating (blue). 
Part (b) is a profile of the grating and nearby layers, showing the reduced pitch on the distal end of the waveguide.
The main panel of (c) shows the fraction of the full solid angle subtended by the grating (dotted blue line) as a function of optic length $l$, for $h=50$\mic\!, $a=62$\mic\!, and $b=50$\mic\!.  This is the fraction of ion light, emitted spherically symmetrically, that would strike the grating. The green line shows the fraction of the light emitted from an ion that would strike the grating and be delivered into the associated waveguide in the single, horizontally polarized transverse mode at the emission wavelength, based on a simulation. The inset in part (c) shows the trade-off between the radial trap frequency (solid line, a measure of trap strength), and solid angle exposure of the ion to the grating (dashed line), when the trap geometry is varied but the ion height and voltage are held fixed.  These latter constraints result in the RF electrode width and trap frequency going to zero as $a$ approaches 100\mic\!. 
}
\label{fig:mainConcept}
\end{figure}

One of the most promising approaches for monolithic collection optics is based on grating couplers \cite{mekis:2011,mehta:2017,zhao:2020}, which can couple light emitted from an ion into single- or multi-mode waveguides via diffractive and taper structures (see Fig.~\ref{fig:mainConcept}).  
A grating coupler used for collection can be straightforwardly optimized based on its performance as an out-coupler, due to the absence of elements which break time reversal symmetry.  
A longitudinal focusing effect can be produced by varying the local grating pitch $\Lambda$ based on the desired emission angle and the relative position of the teeth along the grating length. To minimize the number of diffracted orders, $\Lambda$ is set to the smallest pitch that will diffract normal to the plane of the grating coupler, which is $\Lambda = \lambda/2$ directly beneath the ion.  The guided wavelength is $\lambda = \lambda_0/n_{\rm eff}$, where $\lambda_0$ is the free-space wavelength and $n_{\rm eff}$ is the effective propagation constant of the grating. For forward diffraction angles away from normal, such as for teeth on the end of the grating between the ion and the collection waveguide, $\Lambda > \lambda/2$, while at backward diffraction angles, for teeth on the grating end opposite the waveguide, $\Lambda < \lambda/2$ (see Fig.~\ref{fig:mainConcept}(b)).  Focusing in the lateral dimension can be achieved by curving the grating teeth, analogously to a lens. 

Fabrication constraints limit the minimum resolvable grating pitch, thereby limiting the length of a grating coupler on the backward-emitting end. For example, using 193 nm photolithography in many wafer-scale research-fabrication facilities, lines and spaces can be limited to approximately 120 nm in width, which corresponds to a minimum pitch of about 240 nm. 
Grating couplers designed to collect 400 nm light are therefore unable to extend outward from the waveguide much farther than the projected location of the ion in the PIC plane without resorting to the use of higher resolution lithography. The grating length is also limited on the forward-emitting end of the grating. As teeth are placed farther from the ion, the diffraction angle (with respect to normal) increases. 
Coupling efficiency decreases at steeper diffraction angles and transmission across the air-waveguide surface can be limited by increased Fresnel reflection.  Additionally, for gratings with constant scattering strength as a function of position, increased grating length increases the probability that ion emission is scattered back out of the grating before reaching the waveguide. There is therefore an optimum grating length that balances these trade offs. 

For gratings that have a constant diffraction efficiency throughout, the emission of the grating will have exponentially decaying intensity along its length (proportional to the intensity within the grating). This limits the maximum effective length of the grating, as well as reduces the coupling efficiency to the ion due to reduced mode-matching between the ion emission and grating coupler diffraction. This can be addressed by spatially varying the diffraction strength (apodization) along the length of the grating, which can be effected with a variety of techniques, such as varied etch depths \cite{li:2013,bozzola:2015}, digital patterning of subwavelength features \cite{li:2020}, and/or reduced duty cycles ~\cite{mehta:2020}. Each of these apodization techniques can be limited by fabrication challenges with accurate partial etching or lithographically resolving small feature sizes, e.g. with electron-beam lithography. 

Coupling efficiency is also affected by the vertical structure of a grating.  Single-layer waveguide grating couplers with fully etched grating teeth diffract upward and downward with equal efficiency (assuming a vertically symmetric etch profile), limiting the collection/transmission efficiency to 50\%.  This limitation can be mitigated using dual-waveguide layer gratings with patterning that breaks the symmetry and increases the grating diffraction in the upward direction~\cite{wade:2014}, as well as with reflectors below the grating~\cite{romero:2013}.

\subsubsection{Space constraint trade-offs for optical and trapping elements}

A fundamental limitation on the amount of light that can be collected by a grating between the RF electrodes in a surface trap is the solid angle exposure to the ion. Since the position of the RF null, and therefore ion height ($h$, from Fig.~\ref{fig:mainConcept}(a)), scales with the separation $a$ and width $b$ of the RF electrodes for a simple surface trap with a ground plane under the grating, increasing the gap and other dimensions proportionally will not change the solid angle or amount of light collection.  An absolute upper bound of $\pi$ steradians (corresponding to a fractional exposure of 25\%) can be calculated using an infinitely long collection surface and infinitely thin RF electrodes that maximize the gap size ($a=2h$), though this corresponds to an unrealistic geometry that would require extremely high voltages to trap an ion and consume long axial sections of the trap.  If a geometry is chosen to maximize trap strength (i.e. ion motional frequency) for a given voltage, the fractional exposure for an infinitely long space between the RF electrodes is reduced to 12.5\%.  For $h=50$\mic, this corresponds to $a=41$\mic and $b=100$\mic.

The inset in Fig.~\ref{fig:mainConcept}(c) shows this trade-off when the length of the collection optic is also reasonably constrained. The red, solid curve is a calculation of the normalized radial motional frequency for a trap with fixed voltage, drive frequency, and an ion height of $h=50$\mic above the electrodes. The ion height is preserved for different RF gaps by varying the RF electrode width.  The blue, dashed curve shows the solid angle exposure for a grating with length $l=100$\mic and gap $a=\sqrt{b^2+4h^2}-b$ (derived from~\cite{house:2008}). The plot shows how higher collection efficiencies can be achieved for larger gaps $a$, but at the cost of reduced trap strength for $a>41$\mic\! (given a fixed voltage).

For the purposes of identifying reasonable performance parameters, the rest of the analysis in this section assumes an RF gap of $a=62$\mic as a compromise, corresponding to an RF electrode width of $b=50$\mic\!.
With a length $l=100$\mic this geometry achieves a total exposure fraction of 12.2\% (Fig.~\ref{fig:mainConcept}(c), main panel).

While the analysis in this paper assumes opaque electrodes, it should be noted that these geometric collection limits may be surpassed by using transparent conducting electrodes~\cite{Eltony_2013}. Such an approach would allow a larger solid angle for collection, though there is a trade-off between electrode conductivity and optical transmission that could lead to excess RF dissipation or reduced collection efficiency, at the respective limits of thin and thick transparent electrodes.  

\subsubsection{Fabrication and technical limits}
\label{sec:fablimits}
The analysis above establishes a reasonable upper bound for total collection efficiency based on trap geometry and solid angle exposure, but several technical challenges can reduce it below this value. Most significantly, gratings are not perfectly efficient at coupling light into single-mode waveguides due to limits in diffraction efficiency, lithographic limits, fabrication imperfections, and optical absorption in the grating material.  Many of these shortcomings can be addressed via a combination of more advanced lithographic techniques and 
lower loss waveguide and cladding materials.
While alignment of the ion and optic is much more robust using monolithic integration when compared to free-space techniques, fabrication and material variations can nonetheless also lead to deviations from the designed spatial mode.  

Fig.~\ref{fig:mainConcept}(c) shows the cost of diffraction efficiency by comparing the ideal solid angle collection efficiency to simulations of a single-layer \sini chirped-grating guiding light into a waveguide with the same 62\mic width.  The model simulated here is a single layer, non-apodized, variable pitch grating made from 50 nm thick \sini\!, and uses a simulation wavelength of 493~nm (cf. Table~\ref{tab:species}).
First, a two-dimensional finite-difference time-domain (FDTD) simulation is used to determine the spacings between grating teeth for focusing light at the ion location in the 2D simulation plane, given a maximum grating length (75\mic in this case).
Starting with this design, curved gratings that focus the light in the dimension perpendicular to the 2D plane are generated using a semi-analytic approach that maintains constant optical path lengths from each point on the teeth to the ion.  This approach accounts for the effective index of the light at the grating, including the prior grating teeth the light passes through.  Then three-dimensional FDTD simulations with different grating exposure lengths are used to extract the corresponding collection efficiencies.  
The simulations placed the grating 5\mic below the trap electrodes (assumed to be thin), as shown in part (b) of the figure. The simulation includes the 5\mic of oxide above the grating and below the electrodes.
Simulation points are only shown to a maximum length of 55\mic because longer lengths have reduced collection efficiency due to scattering, a result of the non-apodized gratings simulated here.

The ion was modeled as a dipole source oriented with its axis perpendicular to the trap axis and parallel to the surface.  For simplicity, only $\pi$-emission with the polarization vector oriented along the axis of the dipole source was modeled for this grating.  The location of the dipole source was optimized for the longest grating length using three-dimensional FDTD simulations.  For shorter grating lengths, the dipole location was kept constant, while the length of the grating was reduced symmetrically from both ends. It should be noted that each design instance is not strictly optimal, even under the design restrictions, as that would have required separate two-dimensional and three-dimensional optimizations for each length. A key factor limiting grating collection efficiency in this simple example is the lack of grating directionality afforded in the single-layer, fully-etched grating (halving the maximum possible collection efficiency). This can be mitigated by using multiply-patterned grating layers~\cite{mak:2018} and/or reflection techniques~\cite{romero:2013}; however, in the latter case the vertical separation must be controlled precisely to ensure constructive interference.
Additional refinements such as optimizing waveguide-core thickness, grating apodization~\cite{mehta:2017}, case-specific  optimizations (as opposed to a generic example), and application of more sophisticated design optimizations, such as inverse design~\cite{hammond:2022}, would improve upon this example illustration.

For gratings integrated into surface-electrode ion traps, the material deposition, patterning, and etching processes that define the waveguides and gratings occur within a many-step process that can include multiple metal levels~\cite{niffenegger:2020,mehta:2020,blain:2021} between which the waveguides and gratings are placed.  They therefore must be compatible with the fabrication and geometry of the rest of the trap. For example, the tapers~\cite{Morgan:23} that are needed to match wide collection gratings must avoid the electrical vias, loading holes, and other integrated elements in a typical chip-scale trap~\cite{blain:2021}.
Other considerations include the compatibility of waveguide-core-material deposition temperatures with metal layers, planarization tolerances, and achievable cladding oxide thickness.

Ion behavior is also an important consideration.  As integrated waveguides and gratings have been developed and used to deliver light to ions in ion-trapping structures, one of the primary concerns is dielectric charging that can lead to stray background fields and motional heating. The former effect may be exacerbated when light is delivered through the grating~\cite{ivory:2021}, but this effect will likely be negligible for collection gratings due to the relatively insignificant intensity at the grating of light emitted by the ion.  Nevertheless, the presence of control laser beams near the additional dielectrics included in monolithic designs must be considered.

\subsubsection{Gratings for polarization-based entanglement}
\label{sec:gratings_polarization_PME}

\begin{figure}[tbhp]
\includegraphics[width=\textwidth]{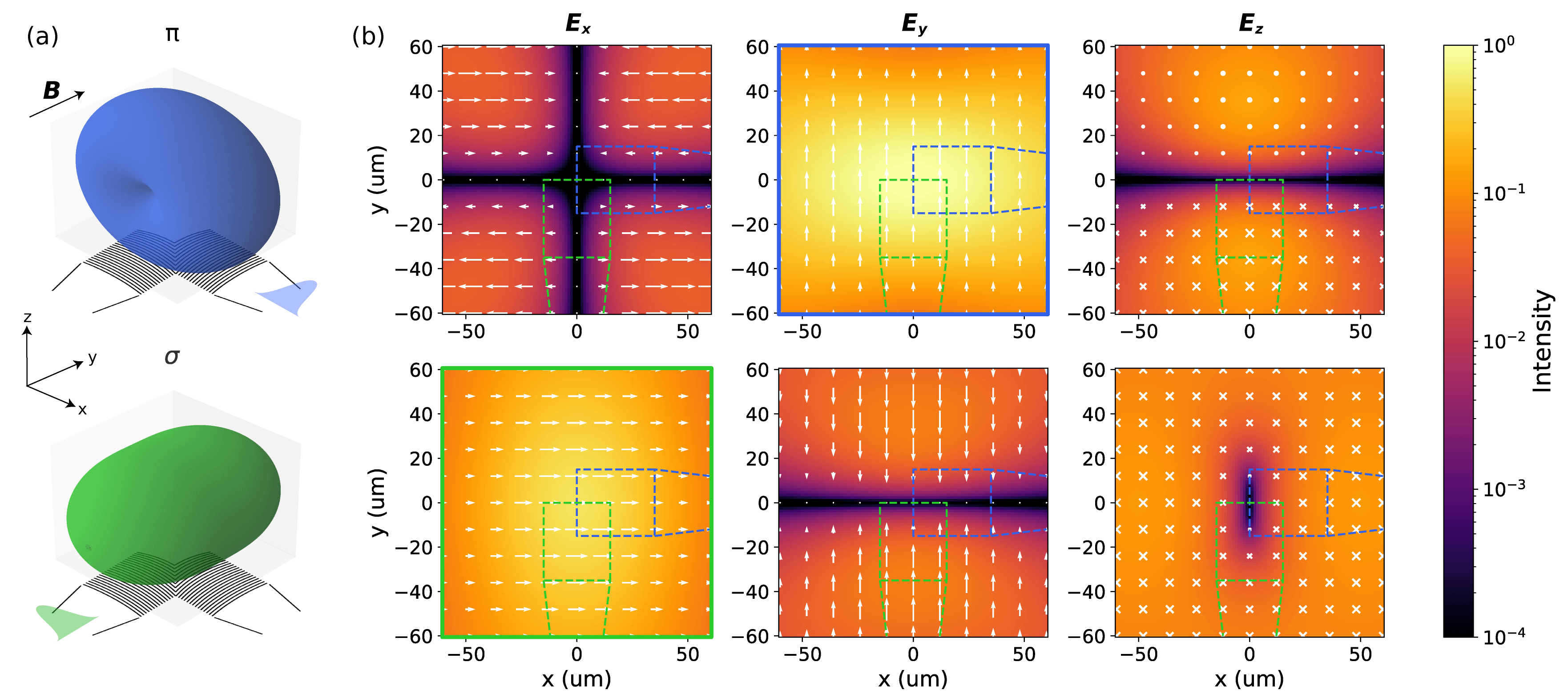}
\caption{Polarization and intensity of photons emitted via $\sigma$ and $\pi$ transitions in a trapped ion.  (a) 3D-renderings of the emission profiles from $\pi$ and $\sigma$ transitions from an ion trapped 50 $\mu$m above the surface of a PIC with collection grating couplers. The relative probability of emission in a given direction is indicated by the radial distance from the origin (ion location) to the surface. (b) The intensity of light emitted from the ion, in arbitrary units, of the $\pi$ and $\sigma$ transitions is illustrated in their respective rows, mapped onto the vector components of the PIC reference frame. The electric field vector components are superimposed in white. Dashed outlines of grating coupler apertures and feeding waveguides are also superimposed in the color corresponding to the ion transition that each grating coupler is designed to collect. As described in the body, the $\pi$ emission is dominantly collected into the TE0 mode of the waveguide running parallel to the $x$ axis (blue), and the $\sigma$ emission is collected into the TE0 mode of the waveguide running parallel to the $y$ axis (green). The dominant field component contributions from each transition are outlined in the color corresponding to the intended collection grating.}  
\label{fig:polarization_PME_with_PICS}
\end{figure}

Due to the highly polarization-dependent behavior of grating couplers, photon collection for polarization-based PME protocols requires additional consideration compared with the other PME protocols. Fig.~\ref{fig:polarization_PME_with_PICS} illustrates a grating coupler configuration suitable for an integrated approach to polarization-based PME analogous to the approach taken by~\cite{ballance:2020}. Here we define the quantization axis to be aligned with the $y$ axis, as depicted by $\mathbf{B}$ in the figure. We also describe polarization modes in the waveguides as quasi-transverse-electric (TE, linearly polarized in the chip plane) and quasi-transverse-magnetic (TM, linearly polarized perpendicular to the chip plane), with an optional numerical index indicating a particular transverse spatial mode (e.g. TE0 for the fundamental TE mode).  As highlighted in the blue boundary, the $\pi$ transition dominantly radiates energy into the $\mathbf{E}_y$ vector-component, which is collected into the TE0 mode of the waveguide running parallel to the $x$ axis. The other vector components of the $\pi$ emission are not collected into the $x$-aligned waveguide: the $\mathbf{E}_z$ vector-component (upper-right) could be collected into the TM mode, but is suppressed by symmetry, and the $\mathbf{E}_x$ vector-component is not supported by the waveguide as it runs parallel to the propagation direction. Conversely, as highlighted in green, the $\mathbf{E}_x$ vector-component of the $\sigma$ transition is collected into the TE0 mode of the $y$-aligned waveguide, illustrated in green. In each waveguide, a small fraction of the orthogonal state is undesirably collected into the TM0 mode via the $\mathbf{E}_z$ vector-components. 

Simulations of the collection of light emitted from an ion into the combination gratings depicted in Fig.~\ref{fig:polarization_PME_with_PICS}  enable us to quantify the expected cross-talk between the two different emitted modes (corresponding to the two different transitions in the ion).  There are two different types of cross-talk for the polarization-based collection.  The first, of primary concern, is the situation in which light from one emission mode of the ion (such as $\pi$) gets into the waveguide mode of the collection grating for the orthogonal emission mode (such as $\sigma$, in this case), specifically into the waveguide mode with the same TE polarization.  This kind of cross-talk is highly suppressed by the symmetry of grating design and positioning relative to the ion location in combination with the anti-symmetry of the emitted light phase as described above.  Our simulations suggest this cross-talk is consistent with zero (we see ${\sim}100$~dB suppression, which is near the numerical limit of the simulations).

The second type of cross-talk we consider is light from the undesired emission mode getting into a different polarization mode of the collection-grating waveguide, i.e. light from the $\sigma$ or $\pi$ emission mode that gets into the TM mode of the waveguide meant to collect the opposite emission mode.  The symmetry considerations described above do not hold for our chosen grating geometry, so there will be non-negligible acceptance of cross-talk in this case.    This type of cross-talk depends significantly on the overlap of the free-space projections of the TE and TM modes at the ion location, which in turn depend on the wavelength and grating details.  In simulation, we see cross-talk of this type ranging from $-22$~dB up to $-5$~dB in the test cases we have explored at two different wavelengths.  This optical cross-talk directly impacts Bell-state fidelity, and if not mitigated, could lead to errors from <1\% to up to ${\sim}30$\%, depending on wavelength and grating details.  However, in the cases where the potential for this cross-talk is high, further filtering of the unwanted TM light can be provided by straightforward bent-waveguide structures that preferentially attenuate light in the TM mode (extinction ratios of >30~dB have been demonstrated~\cite{Zafar:18}), bringing cross-talk-based infidelities to well below the thresholds for error-correcting codes and entanglement distillation protocols. 

From the 3D-rendering of the $\pi$ and $\sigma$ emission patterns, one can perceive that the $\pi$ emission radiates toward the PIC with greater probability, such that the net effect is that $\pi$ emission is twice as likely to be collected by the PIC-waveguide as $\sigma$ emission. This is exactly offset by the fact that the $\sigma$ transition is twice as likely as the $\pi$ transition due to the relevant Clebsch-Gordan factors for a $J=1/2$ to $J=1/2$ transition in a zero-nuclear-spin ion~\cite{ballance:2020}, so there is equal probability of receiving a photon in the blue $x$-aligned waveguide and the green $y$-aligned waveguide.

\subsection{Waveguides and routing}
\label{sec:routing}

Light is collected from the ion using a grating coupler and is then diffracted into a waveguide to route the photons to other locations on the chip and to photonic elements (splitters, combiners, detectors) that make up the circuit. The waveguides need to have low loss at wavelengths of interest to avoid reducing the entanglement rate.  Furthermore, to avoid reduction in Bell-state fidelity, they must have low cross-talk between modes, i.e.\ between neighboring waveguide modes, different spatial modes in the same waveguide, or different frequency modes, depending on PME protocol.  A further requirement is for tight and low-loss waveguide bends that work equivalently for the different photon encodings involved in the routing, in particular for the different waveguide modes needed for the polarization-based entanglement protocol. Recent work in visible light integrated photonics for atomic systems has achieved sufficiently low propagation and bending loss so as to be insignificant for typical envisioned circuit sizes~\cite{west_2019,sorace-agaskar_2019, mcguinness:2022}.

\subsection{Splitters and combiners}
\label{sec:splitters}

The key aspect of each of the PME protocols considered here is the photon interference that destroys the which-path information and allows the photonic degrees of freedom to be traced out to project the ions into an entangled state. In free space this is achieved with the Hong-Ou-Mandel effect at a (non-polarizing) beamsplitter~\cite{HOM_1987}. Mode-selective elements can be utilized in the Bell-state analyzer in order to detect both the $| \Psi^{\pm} \rangle$ states, and thereby maximize the entangled-state production rate.  In this subsection we describe protocol-specific components that can perform these operations in the integrated case.

\subsubsection{Standard splitters/combiners}
\label{sec:normal_splitters}

For number and time-bin protocols, which-path information can be destroyed through use of a simple 4-port splitter (combiner). Such splitters should typically be functional for the frequency based protocols as well, since most standard photonic 4-port splitters will treat frequencies separated by a few tens (or even a few hundreds) of gigahertz  equally.

There are three main types of 4-port splitters in use in integrated photonic circuits, each with their own advantages and disadvantages: directional couplers, adiabatic couplers, and multi-mode interferometers. Directional couplers benefit from simplicity in design and from the fact that symmetry ensures that the two input waveguides are nominally equivalent. However, of the three options, directional couplers are the most sensitive to optical frequency and fabrication variations owing to their dependence on the exact effective refractive index of the waveguides.

Adiabatic couplers benefit from high tolerance to fabrication variations. However, their adiabatic cross-sectional structure requires a larger areal footprint. Additionally, because the two input ports have different dimensions, they may be subject to differential loss or splitting, which can potentially lead to reduction in Bell state fidelity when compared to other splitter designs.

Multi-mode interferometers have the benefits of being compact, symmetric between inputs, and fabrication tolerant. Their main drawback are their abrupt waveguide-dimension discontinuities, which are potential sources of scattering loss and back-reflection. However, various techniques have been developed to minimize losses and reflections~\cite{Pennings_1994,Frishman_2023}, which would be applicable to PME systems.

Imperfection in the Bell-state-analyzer beamsplitting element can lead to infidelities in heralded entangled states through the partial revelation of which-path information. The basic integrated beamsplitters described in this section can likely achieve reflection/transmission imperfections at the 1\% level or below~\cite{Cabanillas:21}.  Since the impact on Bell-state fidelity is quadratic in the difference from a 50:50 splitting ratio~\cite{stephenson_thesis}, we expect the error from these imperfections to be at the ${\sim}1\times 10^{-4}$ level.

\begin{figure}[tbhp]
\centering
\includegraphics[width=0.7 \textwidth]{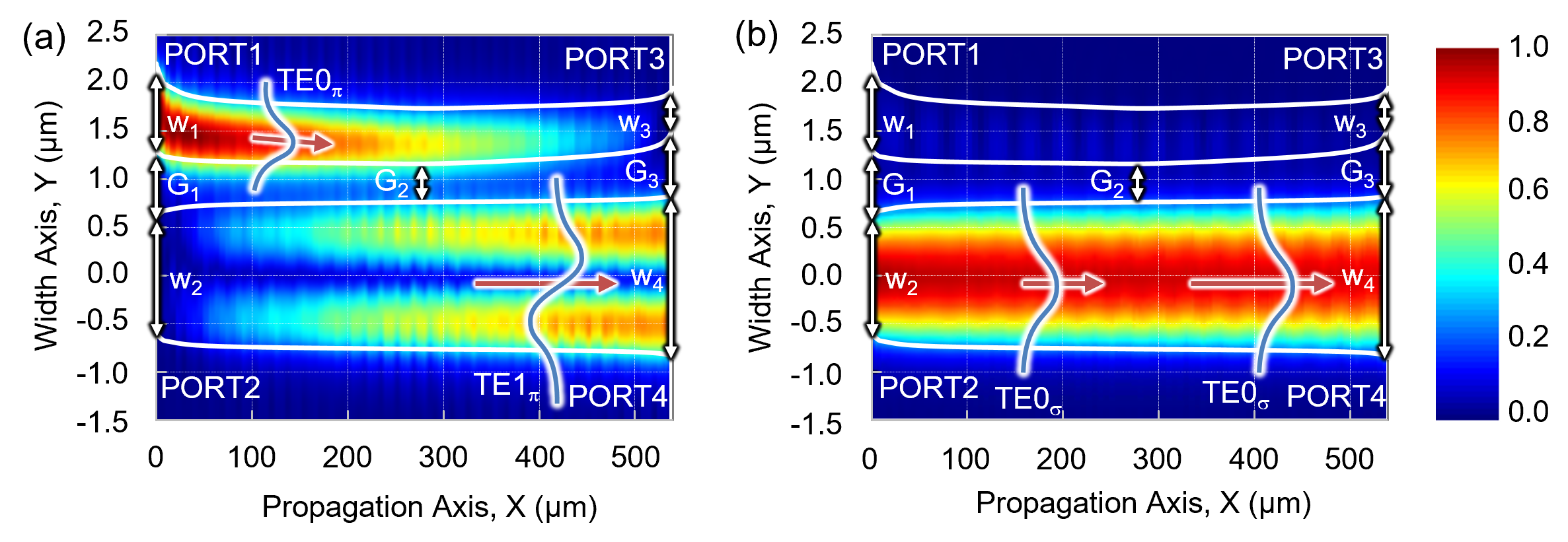}
\caption{Eigenmode expansion simulation of an adiabatic SM-MUX implemented in alumina waveguides for a wavelength of 493 nm. Waveguide edges are outlined in white; color, from blue to red, indicates the relative magnitude of the electric field. Overlay shapes depict relative phase and intensity. (a) TE0-TE1 conversion from port 1 to port 4 (conversion loss $<1$\%), and (b) unconverted TE0 propagation from port 2 to port 4 (transmission loss $<0.1$\%).}
\label{fig:SpatialModeMultiplexer}
\end{figure}

\subsubsection{Integrated optics for polarization-based entanglement}
\label{sec:pol_splitters}
For polarization based protocols that are dependent on the light being emitted into different dipole emission patterns of the ion (e.g.\ $\pi$ or $\sigma$), light can be collected into the TE0 modes of two different waveguides, as described in Sec.~\ref{sec:gratings} and Fig.~\ref{fig:polarization_PME_with_PICS}. Entanglement requires combining the $\pi$ and $\sigma$ light collected by these gratings into a single waveguide while preserving its orthogonal nature. This can be done by mapping the $\pi\-$ and $\sigma\-$ emissions into orthogonal waveguide modes of a single waveguide using a mode multiplexer (MUX). The photons from both ions must be interfered to erase the which-path information regardless of the photon polarization emitted. This can be accomplished by a mode-agnostic splitter/combiner (MAS). To herald an additional Bell state the spatial modes must be de-multiplexed and sent to different detectors, which can be accomplished using a MUX in an inverted orientation. Designs for exemplary MUX and MAS are described below.

\vspace{0.6cm}
\textit{Spatial mode multiplexers}
\vspace{0.5cm}

The required MUX may utilize either orthogonal polarizations (TE, TM) or higher-order spatial modes (TE0, TE1) to preserve orthogonality. Conversion between TE and TM modes can be technologically challenging because TE-TM conversion often requires waveguide widths on the order of the waveguide thickness, which can approach lithographic limits. To avoid this challenge, and to address the aforementioned fabrication sensitivity of directional couplers, we describe in Fig.~\ref{fig:SpatialModeMultiplexer} a spatial-mode MUX (SM-MUX) implemented as an adiabatic coupler; photons entering the two input ports are multiplexed onto different spatial modes of one of the output ports.

The adiabatic coupler is comprised of two waveguides of varying width that are separated by a variable gap $G$, together forming a twin waveguide~\cite{Wang:15,Tambasco:23}. As shown in Fig.~\ref{fig:SpatialModeMultiplexer}(a), a TE0 mode propagating from left to right at port~1 will couple to the TE1 mode of the lower wide multi-mode waveguide, as it is `forced' out of the upper waveguide by the monotonically decreasing waveguide width ($w_3 < w_1$). Selection of appropriate width and gap parameters to realize adiabatic conversion is made by assessing the spatial modes of the upper and lower waveguides in isolation (isolated cases), and together as a twin waveguide (combined case).  At the center of the device, where the gap is at a minimum $G_2$, the widths should be selected such that the propagation constant of the TE0 mode in the upper guide matches the propagation constant of the TE1 of the lower waveguide. Then widths and gaps at the input (left) and output (right) sides should be selected such that select modes of the combined twin waveguide have high overlap with the desired waveguide modes when assessed in isolation. With width and gap parameters suitably selected, the rate of change along the length the coupler can be analytically optimized following methods similar to Refs.~\cite{Siriani:21, Tambasco:23} such that scattering losses are below a desired design criterion (e.g. total scattering losses $\delta< 1$\%). The overall length $L$ scales with the allowed design loss as $L \propto \sqrt{(1/\delta)}$~\cite{Siriani:21}. Overall length for a given loss is minimized when the minimum gap $G_2$ is set to the minimum feature size suitable for good yield in a fabricated device.

Though described here in the `forward’ multiplexing direction, the structure is reciprocal and the same structure will demultiplex TE0 and TE1 modes when operated in reverse. 

\vspace{0.6cm}
\textit{Mode-agnostic splitters}
\vspace{0.5cm}

After mapping the $\pi$- and $\sigma$-emission onto two spatial modes in a common waveguide, a $2\times2$ coupler is needed to combine signals from two collection sites and mix them equally onto two separate waveguides. Many methods exist to split the fundamental mode of an input waveguide onto two outputs, but for this application the same coupler must work equally well for both the TE0 and TE1 modes, and must also preserve the spatial mode mapping at the output. We therefore choose a $2\times2$ multimode-interference splitter architecture for this purpose, due to its inherent self-imaging properties~\cite{soldano_1995} and simplicity of design.

\begin{figure}[tbhp]
\centering
\includegraphics[width=0.6\textwidth]{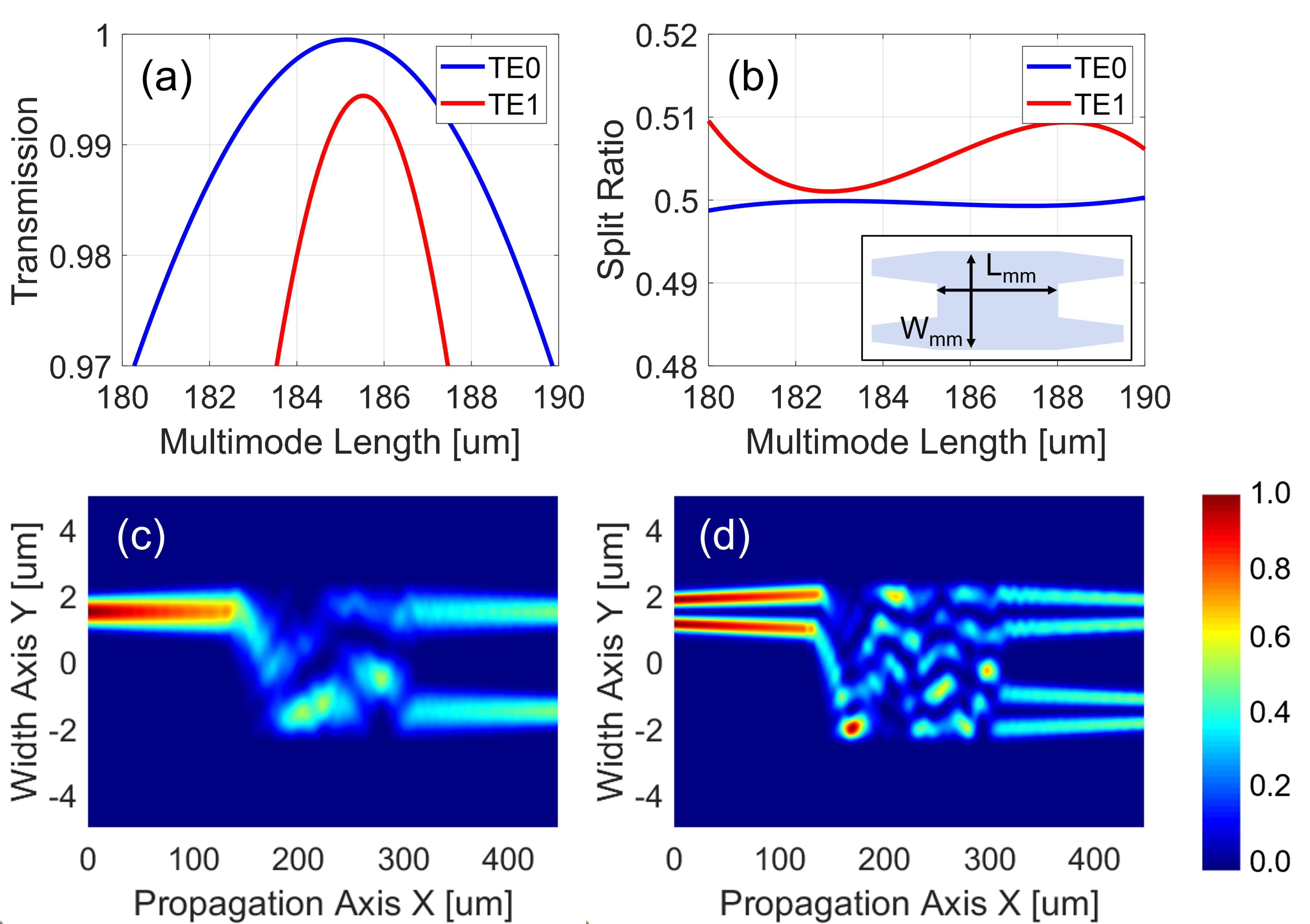}
\caption{Simulated performance metrics of a mode-agnostic splitter designed for 422 nm wavelength light. (a) Transmission and (b) splitting ratio for TE0 and TE1 mode inputs, as a function of the length of the multimode section. The inset in (b) shows a schematic of the device geometry, with primary design parameters denoted. Simulated field intensities are shown for (c) TE0 and (d) TE1 mode inputs.  The color, from dark blue to dark red, indicates the relative amplitude from low to high, respectively.} 
\label{fig:MASfig}
\end{figure}

A schematic illustration of a MAS consisting of linear tapers connected to a wide multimode waveguide is shown in Fig.~\ref{fig:MASfig} [inset of part (b)]. There is a trade-off with respect to the width of the multimode section -- a wider waveguide supporting more modes will present a more complete basis set for the input field at the feed waveguide to project onto, and will therefore result in lower insertion loss. However, the more spatial modes that are excited, the longer the device needs to be in order to obtain the appropriate phase relationship between all modes for equal splitting at the output. We find that a modest number, nine or ten supported modes, strikes a good balance between these considerations. The device performance was simulated using the eigenmode expansion method, and the results for a device designed at 422 nm wavelength are shown in Fig.~\ref{fig:MASfig}(a,b). It can be seen from the plots that a length of approximately 185~\mic gives good performance in terms of both insertion loss and power balance, for both TE0 and TE1 input modes. Fig.~\ref{fig:MASfig} (c,d) shows the simulated electric field intensity distribution in the optimized device, illustrating the mode-agnostic operation of the design.  The simulation suggests that this beamsplitter also attains a deviation from 50:50 performance smaller than 1\%, leading to potential infidelities from this source of error below $10^{-4}$~\cite{stephenson_thesis}.

\subsubsection{Integrated frequency-selective optics}
\label{sec:freq_components}

Frequency-based entanglement relies on the ability to collect photons separated in frequency by several gigahertz and route them through a PIC in a frequency non-discriminating manner. Due to the relatively small frequency spacing compared to the photon frequency, achieving non-discriminating light collection and routing is straightforward using standard components such as grating couplers, tapers, waveguides, and splitters, which have bandwidths $>$10~THz in the blue. 

The main photonics challenge for implementing optimal frequency-based entanglement using PICs is in designing a frequency-selective element for improved Bell state analysis that is narrowband, is stable over timescales relevant to performing entanglement operations, can be tuned to the relevant atomic transition frequencies without requiring an overly burdensome control system, and is relatively compact allowing for implementation in a scalable architecture. Due to the relatively narrow frequency spacing of the photons of interest, the frequency-selective element likely requires the use of resonant or interferometric filters employing cascaded ring resonators or Mach Zehnder interferometers (MZIs), though a grating-based filter may also be an option. Because of the many trade-offs and considerations involved with designing narrowband filters for closely spaced signals, we will not go into great depth about filter design here, but instead we will refer the reader to the literature on microwave photonic signal processing where such filters have been discussed extensively, such as in Ref.~\cite{Liu:20}.

However, several items are worth noting in our discussion. First, material loss becomes the limiting factor in achieving gigahertz-class narrowband filters at telecom wavelengths, which translates to filters with bandwidths about 4$\times$ larger in the visible. Limitations to filter bandwidth may also be exacerbated due to increased material loss and scattering from visible light waveguides compared to telecom. Second, gigahertz-class filters require long optical path lengths and result in device footprints of approximately a few tenths of a square millimeter for ring resonators, and potentially larger areas for MZIs. Such large devices may place constraints on how widely a frequency-based entanglement PIC could be proliferated in a scaled architecture (see Sec.~\ref{sec:analysis}). Finally, many narrowband filters use some of the transmitted photons in feedback loops that control electronics for active devices such as heaters or modulators. In the case of photon-mediated entanglement, a pilot tone may need to be added to the filter to provide feedback to the control system. This can increase PIC complexity, and great care must be taken to avoid crosstalk between this pilot tone and the photons carrying the entanglement information from trapped ions, as this can lead to Bell-state infidelity.

\subsection{Integrated detectors}
Though photon detection may be performed off chip, integrated detectors are necessary to complete the full-stack integration of PME components. These can be employed to generate remote entanglement even when one ion node is off-chip by timing the excitation pulses so that the photon arrival times at the on-chip interference elements are matched for both the remote and the local node~\cite{krutyanskiy:2023}.  Furthermore, their use can mitigate photon loss associated with off-chip coupling to external detectors, as waveguide-coupled detection has been demonstrated with very low loss~\cite{ferrari:2018}.  

Two types of monolithically integrated single-photon detectors have been demonstrated to date in trapped-ion systems: superconducting nanowire single photon detectors (SNSPDs)~\cite{todaro:2021,hampel:2023} and single photon avalanche diodes (SPADs)~\cite{setzer:2021,reens:2022}. SNSPDs have demonstrated a photon detection efficiency (PDE) of 68\% when integrated with an ion trap and have operated at temperatures of up to 6 K, while trap-integrated SPADs can operate at room temperature and have shown PDE up to 40\% \cite{reens:2022}.  The higher detection efficiency attainable with waveguide coupled SNSPDs could substantially increase entanglement generation rates~\cite{ferrari:2018}. While dark count rates for both detector types would be negligible for the short and deterministic ${\sim}$10~ns excitation windows for photon arrival, minimizing stray photons that may arrive at the detector due to the on-chip photonics is crucial to avoid degradation of entanglement fidelity. 

Benefits of waveguide-coupling the single-photon detectors~\cite{moody:2022} include the ability to remotely place the detectors in a part of the chip where they are less likely to be affected by stray light (typically concentrated near the ion) or large Paul-trap RF fields that can limit performance~\cite{slichter:2017}.
Furthermore, detectors fed by a waveguide could allow filtering of ion fluorescence using other integrated photonic elements to reduce stray light background.

\section{Discussion}
\label{sec:analysis}

The above treatment of photonics components for the various PME protocols suggests that there are several trade-offs to consider when determining which protocols may benefit (or suffer) the most from optics integration for light collection and manipulation.  System-level requirements and cycle-time estimates should be understood for particular architectures in order to make an optimal choice. In this section, we highlight the opportunities and difficulties that integrated PME brings and contrast the protocols with a focus on the ease of photonic integration.

First, we briefly discuss achievable maximum entangling rates.  The integrated photonics components can support attempt rates $>\nobreak1$~MHz, but in practice the attempt rate is limited by the ion transition lifetime, the specific protocol used, and the path length between nodes.  The main difference in $r_{\rm ent}$ between PME with integrated components and with bulk optics will be collection efficiency.  From geometric arguments, the achievable collection efficiency in the integrated case will be similar to the state of the art with high-NA lenses (see Sec.~\ref{sec:gratings}).  While our simulations (Fig.~\ref{fig:mainConcept}(c)) do not optimize all parameters (e.g.\ the grating was not apodized nor was the waveguide thickness optimized), we can expect coupling inefficiencies arising from fabrication limits, chiefly resolution limits and edge-roughness, and the achievable index contrast in realistic materials, to limit the attainable collection efficiency. These reductions would likely result in a lower total collection efficiency compared to highly optimized free-space optics~\cite{carter:2023}.
However, integrated PME elements have a more favorable scaling path that can circumvent this rate limit by multiplexing many ion pairs to achieve a higher overall entangled-pair production rate in a similar trap area.

\subsection{Robustness to differential photon loss}

The four protocols have varying susceptibility to differential photon loss before the point of interference of the two photon-collection channels. The number protocol is particularly vulnerable, since entanglement is heralded by a single photon. Therefore, unbalanced losses lead to a bias towards detecting photons from the less lossy node, skewing the state away from the desired Bell state.  This differential loss can be compensated by tuning the excitation probabilities in each node to achieve $p_{e1} \epsilon_1 =p_{e2} \epsilon_2$~\cite{hermans:2023}, which could be achieved with integrated amplitude modulators \cite{hogle:2023}.  However, this fidelity improvement comes at a substantial cost in $r_{\rm ent}$ if the losses are very unbalanced, e.g.\ when networking between on-chip and remote off-chip nodes.

Because they rely on detecting two photons, the Bell-state fidelity obtained via the other protocols is not affected by unbiased differential loss between channels. It is only affected when the two nodes have different loss rates for each photonic qubit state, i.e.\ polarization, frequency, or time bin, since this skews the final entangled state in much the same way that differential photon loss does for the number protocol.  Unlike the number-protocol case, however, the fidelity loss cannot be straightforwardly compensated by tuning the excitation probability between the photonic qubit states. In an integrated platform, such imbalance is a potential concern for polarization-encoded qubits, where mode conversion and propagation are inherently more dependent on the photon properties used for state encoding.  Moreover, they are differentially subject to edge roughness and other imperfections and likely less tolerant to fabrication variation.  Gigahertz-spaced frequencies or microsecond-split time-bins are likely more robust, as extreme frequency sensitivity or periodically varying differential loss between channels is not expected or observed.

\subsection{Potential spatial mode matching gains}

All protocols rely on spatial mode overlap to eliminate which-path information and allow measurement of the photon degrees of freedom to project the ions into an entangled state. On a free-space beamsplitter, this matching is performed by carefully overlapping the decay paths of the nodes, with typical free-space alignment tolerances leading to a limitation near ${\sim}99\%$ mode overlap. On the other hand, in a PIC the photons are confined to discrete spatial modes and a much higher degree of spatial mode overlap can be attained. Currently the infidelity contribution from imperfect spatial mode matching in free-space PME is at most at the 1.3\% level~\cite{ballance:2020}, suggesting that up to 20\% of the total Bell-state error in state-of-the-art experiments could be eliminated via integrated optics. Moreover, the inherent mode matching in waveguide-based devices increases robustness and stability by obviating the need for careful alignment, and frequent re-alignment, of the individual paths.  This robustness paves the way for scaling to higher numbers of nodes and for large-scale multiplexing within each node to increase entanglement rates.
We note that the properties of integrated beamsplitters for PME as described above (Sec.~\ref{sec:normal_splitters}) disincentivize the use of adiabatic couplers in favor of directional couplers or multi-mode interferometric splitters because the latter are more compact and less likely to cause differential loss or splitting between the paths.

\subsection{Ion species and transition choice}

The considerations for choosing an ion species and transition for a given protocol include ion level structure, transition wavelengths (and hence lithographic limitations), presence of nuclear spin, and ease of conversion to telecom wavelengths (for long-distance remote entanglement generation). As discussed in Sec.~\ref{sec:gratings}, lithographic limits can reduce the efficacy of collection gratings for blue light when compared to red/IR, though advances in photolithography (or electron-beam lithography) may allow for the smaller features needed in this case. Additionally, transmission loss in the standard SiN$_{\rm x}$ or alumina waveguides grows rapidly for wavelengths below ${\sim}$400~nm~\cite{west_2019}.  With transition wavelengths of 455~nm and 493~nm, Ba$^+$ is a good choice based on these considerations.  However, given the many application-dependent criteria that have to be balanced, other species with wavelengths above ${\sim}$400~nm are also viable candidates.

To ease lithographic and transmission limits, one could also use longer wavelength transitions that connect to metastable states in any of these species. However, the branching ratios of these transitions are generally in the $\gamma=2$-$27$\% range (without cavity enhancement) and since the PME rate goes as $\gamma^2$, there must be a useful trade off in a particular application to make this choice tenable.  One such situation may be for remote links, where longer wavelengths are also favored for fiber transmission. However, efficient quantum frequency conversion of single photons from the blue wavelengths suited for PME to the telecom band would obviate this benefit~\cite{saha:2023}.

There are pros and cons to choosing an ion with non-zero nuclear spin and thus hyperfine structure. The frequency protocol typically requires ground-state hyperfine structure to provide the frequency separation of photonic qubit states. Hyperfine structure also allows for ``clock'' states to greatly enhance the qubit coherence time without the need for a co-trapped memory ion. However, the additional atomic levels compared to a zero nuclear spin ion increase the complexity of state preparation and manipulation.

\subsection{Protocol comparison}

The frequency protocol has the potential to achieve a fast entanglement rate with comparatively simple integrated photonic elements.  However, the presence of hyperfine structure complicates state preparation and the typical $\sim$10~GHz hyperfine splitting requires high finesse, resonant structures, with concomitant requirements on path-length stabilization, in order to distinguish between the photon frequencies in a Bell-state analysis (see Sec.~\ref{sec:freq_components}).  There is a penalty of a factor of $\frac{1}{2}$ in PME rate when not making this frequency distinction.  If this reduction in rate due to not incorporating enhanced Bell-state analysis can be tolerated, the frequency protocol may be highly promising for integrated PME.

The time-bin protocol places minimal requirements on the photonic devices and can be implemented using ions without hyperfine structure.  The fidelity for this protocol is limited by the single qubit rotations used to create the initial superposition and for inversion of the qubit manifold during the protocol.  These high-fidelity, single-qubit rotations could be provided via an additional integrated light path (for state-control in this case) for each PME site---utilizing such an architecture could make the time-bin protocol promising via integrated optics.   

The number protocol gains the most from photonic integration because the required optical path-length stability is inherent as long as decay channels are combined on chip, and thus improved over typical free-space implementations. The photonic elements and required level structure are the same as for the time-bin protocol and thus straightforward.  The drawbacks of this protocol are the inherent trade-off between fidelity and rate, the requirement to be near the ground state of motion for high fidelity, and the requirement of balanced photon probability in each channel. These downsides mean that the number protocol is likely not suitable for very high-fidelity, high-rate PME. However, for devices with low collection efficiency, e.g.\ where geometric factors limit collection grating size, but that have high confinement, this protocol allows an increase in entanglement rate over two-photon schemes by a factor of $4\frac{p_e}{\epsilon}$~\cite{humphreys:2018}.

As presented in Sec.~\ref{sec:splitters}, the polarization protocol uses bespoke photonic devices not needed for the other schemes. Assuming spatial-mode multiplexing of polarization modes as described here, the additional complexity of discriminating between photon polarizations in the Bell-state analysis is marginal and therefore worth implementing to achieve a factor of two in entanglement rate. The benefits of this protocol are that the ion manipulation is straightforward and among the fastest of all the protocols we consider. There is also no inherent fidelity limit depending on other operations (such as single qubit rotations). A unique challenge is that more control of achievable device geometry may be required, for example to avoid differential photon loss between modes (polarizations or spatial modes) or imperfect mapping of the emitted-photon polarization to different spatial modes.

\subsection{Multiplexing and scalability} 

As discussed at the beginning of this section, the maximum achievable $r_{\rm ent}$ will likely not be higher for integrated optics compared to traditional bulk optics that are optimized for collection {\it at a single-site}.  When considering multi-site PME to enhance overall rate, however, the size of integrated optics relative to the ion array pitch may provide a considerable advantage.  A bulk lens consumes significant lateral chip area that precludes collection from neighboring sites, while microfabricated PME elements, like those described in this paper, are scaled to the trap size and do not prevent any other sites from being used for PME.
The lower collection efficiency per element in the integrated case could therefore be more than offset by the increased number of sites used for PME.  If these sites were used in a coordinated, multiplexed protocol with continuous excitation of many ion pairs in parallel, heralded entanglement between any particular pair could be transferred to computational ions convenient to all entanglement sites.  Thus the overall system entangled-pair creation rate could be increased.  This comes at the cost of a larger number of ion sites, but such reproducibility is achievable using microfabricated and integrated elements.

The benefits to PME scalability promised by monolithic integration are an array architecture based on a unit-cell that can be multiplexed for increased entanglement rates, a smaller overall system-scale footprint, and simpler and more robust optical alignment.
A unit cell for photon-mediated entanglement would include emission gratings for state preparation and pulsed excitation, collection gratings coupled to waveguides, DC electrodes for ion confinement, and a splitter/combiner and associated detectors.
Using this approach, large numbers of PME cells can be integrated on the same chip, improving overall rates of ion-ion entanglement through multiplexing as explained above.  Such a configuration could reduce ion shuttling and re-cooling requirements in a quantum computing module~\cite{pino:2021}, allowing rapid establishment and distribution of entangled pairs for quantum algorithms. 

In addition to well mode-matched collection gratings, entanglement generation between remote nodes (cf. Fig.~\ref{fig:overview}(b)) will require scalable, low-loss waveguide-to-fiber interfaces, which is an active area of study~\cite{son:2018}. This is an additional consideration for extending processors or quantum networks beyond arrays contained on an individual substrate.  However, it is unlikely to be a limiting factor since coupling loss at the sub-decibel scale and few-decibel scale has been achieved at telecom~\cite{Zhu:23} and visible/infra-red wavelengths~\cite{mehta:2020,Poon:24}, respectively.  A further consideration for connecting multiple substrates using any of these PME protocols is the need for path-length stabilization, depending on the length of the link.  Chip-to-chip links in the same vacuum system may not require active stabilization, but links used in typical long-distance quantum-repeater architectures would likely require it.

\section{Conclusion}
\label{sec:conclusion}
Photon-mediated entanglement is a powerful resource that can enable quantum networking and communication within a trapped-ion system. These photon interactions can also connect different qubit types in a hybrid system to combine the long memory times and high fidelities provided by trapped ions with advantageous properties of other qubits.  We have presented paths towards integrating the optical elements for several photon-mediated entanglement protocols. As with other trapped-ion quantum technologies that have been demonstrated using micro-fabricated, integrated elements, the photonic integration for PME promises benefits including stability, scalability, and manufacturability, while also creating performance challenges to overcome.  Nonetheless, these technologies provide a promising approach to higher overall remote entanglement rates and a potential alternative to the current approach of ion shuttling and local Coulomb interactions to produce ion-ion entanglement in a modular quantum computer on a single ion-trapping chip~\cite{wineland:1998}.



\begin{backmatter}
\bmsection{Funding}
Content in the funding section will be generated entirely from details submitted to Prism. Authors may add placeholder text in the manuscript to assess length, but any text added to this section in the manuscript will be replaced during production and will display official funder names along with any grant numbers provided. If additional details about a funder are required, they may be added to the Acknowledgments, even if this duplicates information in the funding section. See the example below in Acknowledgements. For preprint submissions, please include funder names and grant numbers in the manuscript.

\bmsection{Acknowledgments}
The authors wish to thank Galen Hoffman and Brian DeMarco for useful discussions and thank Gabriel Araneda and David Nadlinger for helpful comments.

This material is based upon work supported by the U.S. Department of Energy, Office of Science, National Quantum Information Science Research Centers, Quantum Systems Accelerator. Additional support is acknowledged from the NSF Quantum Leap Challenge Institute for Hybrid Quantum Architectures (award \#2016136) and through the Q-SEnSE Quantum Leap Challenge Institute (award \#2016244).  This material is based upon work supported by the Department of Energy under Air Force Contract No. FA8702-15-D-0001. Any opinions, findings, conclusions or recommendations expressed in this material are those of the author(s) and do not necessarily reflect the views of the Department of Energy.

Sandia National Laboratories is a multimission laboratory managed and operated by National Technology \& Engineering Solutions of Sandia, LLC, a wholly owned subsidiary of Honeywell International Inc., for the U.S. Department of Energy’s National Nuclear Security Administration under contract DE-NA0003525.  This paper describes objective technical results and analysis. Any subjective views or opinions that might be expressed in the paper do not necessarily represent the views of the U.S. Department of Energy or the United States Government.

\bmsection{Disclosures}
The authors declare no conflicts of interest.






\bmsection{Data Availability Statement}
Data underlying the results presented in this paper are not publicly available at this time but may be obtained from the authors upon reasonable request.

\end{backmatter}

\bibliography{main}

\end{document}